\documentclass[aps,prl, reprint,superscriptaddress,longbibliography,nobibnotes]{revtex4-2}

\usepackage{amsfonts,amscd,mathrsfs,amsmath,amsthm,amssymb}
\usepackage{mathtools}
\usepackage{xparse}
\usepackage{graphicx}
\usepackage{float}
\usepackage{pifont}
\usepackage[dvipsnames]{xcolor}
\usepackage{colortbl}
\usepackage{multirow}
\usepackage{enumerate}   
\usepackage{comment}
\usepackage{booktabs}
\usepackage{cancel}
\usepackage{subcaption}
\usepackage[new]{old-arrows}
\usepackage{circuitikz}

\usepackage{lieart}

\usepackage{listings}
\usepackage{xcolor}

\definecolor{codegreen}{rgb}{0,0.6,0}
\definecolor{codegray}{rgb}{0.2,0.2,0.2}
\definecolor{codepurple}{rgb}{0.58,0,0.82}
\definecolor{backcolour}{rgb}{0.95,0.95,0.92}

\lstdefinestyle{mystyle}{
    backgroundcolor=\color{backcolour},   
    commentstyle=\color{codegreen},
    keywordstyle=[1]\color{magenta},
    keywordstyle=[2]\color{blue},
    otherkeywords = {PerfectGroup,CharacterTable,Irr,Norm,ScalarProduct,gap>},
    morekeywords = [1]{PerfectGroup,CharacterTable,Irr,Norm,ScalarProduct,Degree},
    morekeywords = [2]{gap>,>},
    numberstyle=\tiny\color{codegray},
    stringstyle=\color{codepurple},
    basicstyle=\ttfamily\footnotesize,
    breakatwhitespace=false,         
    breaklines=true,                 
    captionpos=b,                    
    keepspaces=true,                 
    numbers=left,                    
    numbersep=5pt,   
    frame = single, 
    showspaces=false,                
    showstringspaces=false,
    showtabs=false,  
    morecomment=[l][\color{codegreen}]{\#},
    alsoletter=?>,
    tabsize=2
}
\usepackage{physics}
\usepackage{bm}
\usepackage{bbm}
\usepackage{caption}

\PassOptionsToPackage{hyphens}{url}\usepackage{hyperref}

\hypersetup{
    colorlinks=true,
    linkcolor=magenta!80!black,  
    citecolor=magenta!80!black,
    breaklinks=true,
}
\usepackage[capitalise]{cleveref}

\theoremstyle{theorem}

\theoremstyle{definition}

\newtheorem*{conjecture*}{Conjecture}

\AddToHook{cmd/appendix/before}{%
  \setcounter{equation}{0}%
  \renewcommand{\theequation}{\thesection\arabic{equation}}%
  \setcounter{theorem}{0}%
  \setcounter{lemma}{0}%
}

    \newcommand\numberthis{\addtocounter{equation}{1}\tag{\theequation}}

    \newcommand{\R}{\mathbb{R}} 
    \newcommand{\CC}{\mathbb{C}}

    \DeclareMathAlphabet{\mathsfit}{T1}{\sfdefault}{\mddefault}{\sldefault}
    \SetMathAlphabet{\mathsfit}{bold}{T1}{\sfdefault}{\bfdefault}{\sldefault}

    \renewcommand{\F}{\mathsf{F}}

    \renewcommand{\SU}{\mathrm{SU}}

    \renewcommand{\U}{\mathrm{U}}

    \renewcommand{\D}{\mathscr{D}}

\renewcommand{\E}{\mathcal{E}} 

\renewcommand{\F}{\mathcal{F}}

\newcommand{\Inv}{\mathrm{Inv}}

\newcommand{\Sym}{\mathrm{Sym}}

\begin{document}

\title{Classical and Quantum MacWilliams Transforms as Spin Kinematics}

\author{Eric Kubischta}
\email{emk25g@fsu.edu}
\affiliation{Quantum Initiative, Florida State University, Tallahassee, FL 32306}
\affiliation{Department of Mathematics, Florida State University, Tallahassee, FL 32306}

\author{Ian Teixeira}
\email{iteixeira@ucsd.edu}
\affiliation{Department of Mathematics, University of California, San Diego, CA 92093}

\begin{abstract}
Spin is the hidden engine behind the zoo of MacWilliams transforms in weight enumerator theories—not only for qubits and qudits, but even for classical codes. From nothing more than a split into trivial and nontrivial errors, we kinematically derive the MacWilliams transform as a Wigner-$D$ rotation between two canonical bases. Within each classical and quantum theory, changing the length $n$ leaves the rotation untouched: the same element simply reappears at spin $n/2$. And at fixed $n$, changing the rotation axis simply moves between the various classical and quantum theories. 
\end{abstract}

\maketitle

\emph{Introduction.--}
Weight enumerators are one of the miracles of coding theory. The elementary bounds---Hamming, Singleton, Gilbert--Varshamov---are indispensable first cuts, but they are often too blunt to settle sharp parameter questions. The linear-programming revolution in coding theory came from the more delicate constraints satisfied by weight enumerators and their MacWilliams transforms: the weight distribution of a linear code and that of its dual are locked together by a universal matrix, and Delsarte's association-scheme method turns this lock into computable nonexistence tests. This information feeds the MacWilliams identities, Delsarte's linear-programming method, the association-scheme formalism, shadow bounds for self-dual codes, and many of the most effective nonexistence arguments in classical coding theory \cite{MacWilliams1963,MacWilliamsSloane1977,Delsarte1973,ConwaySloane,rainsWE,Shadows}.

The quantum story is parallel.  Shor and Laflamme introduced qubit weight enumerators and a quantum MacWilliams identity relating them, using these with linear programming to constrain quantum codes \cite{ShorWE}.  Rains then introduced shadows and unitary enumerators for qudit codes, tightening the resulting bounds and clarifying several optimality questions \cite{rainsWE,Shadows,RainsPolys}.  Since then, weight enumerators have reappeared in many guises: split and asymmetric enumerators \cite{SplitWE}, complete and double enumerators \cite{WEdoubleandcomplete,CompleteWeightEnumQuantum}, tensor-network enumerators \cite{WEtensor} and, very recently, mixed-dimensional or heterogeneous MacWilliams identities \cite{MixedDimMacWilliams}.

This is a rich theory.  It is also a high barrier to entry: the same object is described by Hamming schemes in the classical case, trace identities in the quantum case, invariant theory for self-dual codes, and tensor contractions for network codes.

The purpose of this Letter is to strip away the realization and expose the kinematics.  We use ``kinematics'' in the physicist's sense: the state space, the preferred frames, and the allowed changes of frame before one chooses a ``Hamiltonian".  Here the state space is the sector space of local error labels; the preferred frames are the sector bases; and the MacWilliams and shadow transforms are changes of frame.  A classical code, quantum projector, stabilizer code, or tensor network supplies a particular vector in this space.  That assignment is a realization.

This separation makes the ordinary MacWilliams transform look almost inevitable.  Ordinary weight enumerators forget which nontrivial local symbol occurred; they remember only whether a site was inactive or active.  One site therefore gives a two-dimensional sector space.  On $n$ identical sites, after forgetting positions and retaining only the number of active sites, the space is not just an arbitrary $(n+1)$-dimensional vector space.  It is the spin-$n/2$ irrep of $\SU(2)$.  The one-site MacWilliams transform is a sector-frame change; the $n$-site transform is its spin-$n/2$ lift.  In orthonormal sector coordinates it is a Wigner-$D$ matrix.  In the counting coordinates preferred by coding theory it becomes the Krawtchouk matrix.

This viewpoint also demystifies qubit shadow enumerators \cite{Shadows}. The global one-site frame consists of a uniform line and a mean-zero line. The uniform line has a distinguished positive unit vector. The mean-zero line has two equally canonical orientations.  Choosing one gives MacWilliams; choosing the other gives the qubit shadow transform.  Thus the shadow is not an extra ad hoc decoration.  It is the remaining sign choice in the same two-dimensional kinematics. 

\begin{table}[t]
    \centering
    \begin{tabular}{lcc}
        \toprule
        Weight-enumerator theory & $N$ & Euler angle $\beta$ \\
        \midrule
        Classical bit
            & $2$
            & $\tfrac{\pi}{2}$ \\[3pt]
        Classical $q$-ary
            & $q$
            & $2\arccos\!\left(\tfrac{1}{\sqrt q}\right)$ \\[6pt]
        Qubit
            & $4$
            & $\tfrac{2\pi}{3}$ \\[3pt]
        Qudit (dimension $D$)
            & $D^2$
            & $2\arccos\!\left(\tfrac{1}{D}\right)$ \\[6pt]
        Rains' unitary
            & $N\to\infty$
            & $\beta\to\pi$ \\
        \bottomrule
    \end{tabular}
    \caption{MacWilliams transforms as a Wigner-$D$ matrix $D^{n/2}(\tfrac{\pi}{2}, \beta, \tfrac{\pi}{2})$ for each of the various theories. Each theory has 1 trivial and $N-1$ non-trivial symbols.}
    \label{tab:macwilliams-spin-family}
\end{table}

\emph{One site kinematics.--}
Consider one site with a finite set $\E$ of local error symbols. These symbols for example may be $q$-ary bit flips in a classical code, or a basis of the local operator algebra in a quantum code.  We assume only a partition into inactive and active sectors,
\begin{equation}
  \E=\E_0\sqcup \E_1,
  \qquad |\E_0|=r,
  \qquad |\E_1|=s,
  \qquad N=r+s .
  \label{eq:partition}
\end{equation}

Here \(\E_0\) denotes the trivial/inactive sector (those errors that effect the site in a trivial way, like the identity operation), while \(\E_1\) is the non-trivial/active sector (those errors that distort the site, e.g., the bit flip). Ordinarily we have $r=1$ and $s = N-1$ because the only trivial error is the identity operation. However, we keep \(r\) and \(s\) arbitrary, since the derivation is no harder and
the same formula also covers potential variants.

Let $f: \E \to \R$ be a function that assigns a real number to each error. The set of all such functions is a real vector space and is denoted $\mathcal{F}(\E) := \R^\E$. The collection of delta functions \(\{\delta_e:e\in\E\}\) is the
standard orthonormal basis of \(\mathcal F(\E)\).

Ordinary weight enumerators do not distinguish symbols inside a fixed sector.  Thus they exhibit the sector-preserving symmetry $G=S_r\times S_s$, which permutes inactive symbols among themselves and active symbols among themselves.  The one-site ordinary sector space is thus the $G$-invariant part of $\F(\E)$ and it is two-dimensional.  A canonical orthonormal basis is
\begin{equation}
  \ket{0}=\frac{1}{\sqrt r}\sum_{e\in\E_0}\delta_e,
  \qquad
  \ket{1}=\frac{1}{\sqrt s}\sum_{e\in\E_1}\delta_e ,
  \label{eq:local-basis}
\end{equation}
which we dub the \textit{local-sector basis}.

There is a second natural frame, arising from the action of $S_N$ that permutes \textit{all} $N$ symbols. We have
\begin{equation}
  \F(\E)\cong {\bf 1}_{S_N}\oplus \mathrm{Std}_{S_N},
\end{equation}
where $\mathrm{Std}_{S_N}$ is the mean-zero subspace $\sum_{e \in \E} f(e) = 0$. Each irreducible contains one $G$-invariant line. 

The uniform vector
\begin{equation}
  \ket{+}=\frac{1}{\sqrt N}\sum_{e\in\E}\delta_e
  =\sqrt\frac{r}{N}\ket{0}+\sqrt\frac{s}{N}\ket{1},
  \label{eq:plus}
\end{equation}
is the canonical basis vector inside ${\bf 1}_{S_N}$.

The $G$-invariant line inside $\mathrm{Std}_{S_N}$ has two canonical choices of orientation. In other words, there is a $\mathbb{Z}_2$ gauge freedom. We denote these by
\begin{align*}
    \ket{-}
    &=
        \sqrt{\tfrac{s}{r N}}
        \sum_{e\in \E_0}\delta_e
        -
        \sqrt{\tfrac{r}{s N}}
        \sum_{e\in \E_1}\delta_e
    = \sqrt{\tfrac{s}{N}}\ket{0}-\sqrt{\tfrac{r}{N}}\ket{1}, \\
    \ket{\sigma} &= - \ket{-}.
\end{align*}
For now, we focus on $\ket{-}$ as it will lead to duality (involution); we call $\{ \ket{+}, \ket{-} \}$ the \textit{global-sector basis}.

The unique change of basis matrix from the local-sector basis to the global-sector basis is
\[
    U
    =
    \frac{1}{\sqrt N}
    \begin{pmatrix}
        \sqrt r & \sqrt s \\
        \sqrt s & -\sqrt r
    \end{pmatrix}. \numberthis
\]
This is the one-site \textit{unitary} MacWilliams transform. It depends only on the sector sizes $r$ and $s$, not on any code, and in this sense it is kinematical.

The matrix $U$ is real and orthogonal.  After a fixed one-site phase convention $\widehat U=
  -iU$, it becomes an element of $\SU(2)$. And, with the $ZXZ$ Euler angle convention, we can express this as a Wigner-$D$ matrix
\begin{align}
  \widehat U&=D^{1/2}\left(\frac{\pi}{2},\beta,\frac{\pi}{2}\right), \qquad \beta = 2 \arccos(\sqrt{\tfrac{r}{N}}).
\end{align}

Everything so far lives in the orthonormal sector basis, where the transform
is manifestly unitary. Coding theory, however, does not count in an orthonormal
basis: a weight-enumerator coefficient is a \emph{number of symbols}, and that
number lives most naturally in the unnormalized basis of sector indicators
\begin{equation}
  \mathbf{x} = \sum_{e\in\E_0}\delta_e = \sqrt r\,\ket 0,
  \qquad
  \mathbf{y} = \sum_{e\in\E_1}\delta_e = \sqrt s\,\ket 1 .
\end{equation}
Passing from the local-sector basis $\{\ket0,\ket1\}$ to the \textit{local-counting basis} $\{\mathbf x,\mathbf y\}$ is the diagonal
rescaling $\Delta=\mathrm{diag}(\sqrt r,\sqrt s)$.  The MacWilliams matrix is then the \emph{same} map $U$, only expressed in these
counting coordinates. Conjugating by the rescaling and clearing the overall
$1/\sqrt N$ so the entries come out integral gives, in the ordinary case
$r=1,\ s=N-1$,
\begin{equation}
  M = \sqrt{N}\,\Delta\,U\,\Delta^{-1}
  = \begin{pmatrix} 1 & 1 \\ N-1 & -1 \end{pmatrix}.
\end{equation}
Thus the familiar one-site MacWilliams matrix is the counting-coordinate
version of the orthonormal basis change between the local-sector basis and 
the global-sector basis. This is the key point: the MacWilliams transform 
is unitary. The familiar matrix $M$ only looks non-unitary because 
coding theorists use counting coordinates rather than orthonormal ones. 
Change coordinates and unitarity is restored.

\emph{Many site kinematics.--}
For $n$ identical sites, ordinary weight enumerator theories forget location and remember only the number $k$ of active sites.  Equivalently, it takes the $n$-th symmetric power of the one-site $G$-invariant sector space.

The $n$-site unitary MacWilliams transform $U^{(n)}$ is obtained by applying the one-site transform $U$ to each site and restricting to the symmetric subspace. 
Consequently
\begin{equation}
  \widehat U^{(n)}=\Sym^n\left(\widehat U^{(1)}\right)
  =D^{n/2}\left(\frac{\pi}{2},\beta,\frac{\pi}{2}\right).
  \label{eq:wigner-n}
\end{equation}
This is the main point.  The ordinary $n$-site unitary MacWilliams transform $ U^{(n)} = i^n \, \widehat{U}^{(n)}$ is (up to a phase) a Wigner-$D$ matrix in the spin-$n/2$ representation:
\[
    U^{(n)} = i^n D^{n/2}\left(\frac{\pi}{2},\beta,\frac{\pi}{2}\right). \numberthis
\]

Just as we did for one site, we can switch to counting coordinates. Suppose we have a length \(n\) error string where \(k\) of the sites are
active and \(n-k\) of the sites are trivial.  The number of such strings is $\binom{n}{k} r^{n-k}s^k$. Thus the \(n\)-site sector-size diagonal matrix is $\Delta^{(n)}$ whose $k$-th entry is $\sqrt{\binom{n}{k} r^{n-k}s^k}$. Then as before, if we conjugate the unitary MacWilliams matrix $U^{(n)}$ by this diagonal rescaling and clear the overall $1/N^{n/2}$ coefficient so again the entries come out integral, we get the standard $r=1,s=N-1$ MacWilliams transform
\[
    M^{(n)} = N^{n/2} \Delta^{(n)} U^{(n)} {\Delta^{(n)}}^{-1}. \numberthis
\]
In the ordinary $r=1, s=N-1$ case, the matrix elements are given by the Krawtchouk formula (which we derive in the appendix)
\begin{equation}
  M^{(n)}_{j,k}=\sum_{\ell=0}^{j}(-1)^\ell
  \binom{k}{\ell}\binom{n-k}{j-\ell}\left( N-1 \right)^{j-\ell} .
  \label{eq:krawtchouk}
\end{equation}

\emph{Realizations.--} So far, from first principles, we have derived the unitary MacWilliams transform as the unique change of basis matrix between two canonical orthonormal bases. Then we have shown how the standard MacWilliams transform is its expression in counting coordinates. This has all been pure kinematics and shows how the MacWilliams transform exists outside of a specific \textit{realization}, that is, outside of how we actually assign numbers to each error. 

Let $\widetilde{A}$ be the coefficients with respect to the local-sector basis and let $\widetilde{B}$ be the coefficients with respect to the global-sector basis so that $\widetilde{B} = U^{(n)} \widetilde{A}$. Similarly let $A$ be the coefficients with respect to the local-counting basis and let $B$ be the coefficients with respect to the global-counting basis so that $B = M^{(n)} A$.

For a linear classical $q$-ary code $C\subseteq\mathbb F_q^n$, the coefficient $A_k$ is chosen to be the number of codewords of weight-$k$ and $B_k$ is chosen to be the number of dual codewords of weight-$k$ \cite{MacWilliamsSloane1977} (both up to a normalization by the total number of codewords). They are related by the MacWilliams transformation \cite{MacWilliams1963} that acts on the counting coordinates as
\begin{align*}
    \mathbf{x} &\to \mathbf{x}+(q-1)\mathbf{y} \\
    \mathbf{y} &\to \mathbf{x}-\mathbf{y}.
\end{align*}
This corresponds to the matrix
\[
    \mqty(1 & 1 \\q-1 & -1),
\]
which is exactly the matrix we derived in \cref{eq:krawtchouk} for $r=1$, $s=q-1$, $N=q$ (and thus justifies us calling it the MacWilliams matrix). 

For a quantum code with projector $P$ of rank $K$ in $n$-many qudits of local dimension $D$, choose a one-site error basis. There are $N=D^2$ local error symbols: the identity and $D^2-1$ active operators. The Shor--Laflamme enumerators are, up to a global normalization,
\begin{align}
  A_j&=\frac{1}{K D^n}\sum_{\mathrm{wt}(E)=j}|\Tr(PE)|^2,\nonumber\\
  B_j&=\frac{1}{K}\sum_{\mathrm{wt}(E)=j}\Tr( PEP E^\dagger) .
\end{align}
These are related by the MacWilliams transformation
\begin{align*}
    \mathbf{x} &\to \mathbf{x}+(D^2-1)\mathbf{y} \\
    \mathbf{y} &\to \mathbf{x}-\mathbf{y}.
\end{align*}
This corresponds to the matrix
\[
    \mqty(1 & 1 \\D^2-1 & -1),
\]
which is exactly the matrix from \cref{eq:krawtchouk} derived before for $r=1$, $s=D^2-1$, $N=D^2$.

So we see that our kinematical theory has for $r=1$, $s=N-1$ produced an infinite family of matrices:
\[
    U = \frac{1}{\sqrt{N}}\smqty(1 & \sqrt{N-1} \\ \sqrt{N-1} & -1), \qquad M = \smqty(1 & 1 \\ N-1 & -1)
\]
which reproduce the unitary and Krawtchouk MacWilliams transforms for bits, $q$-ary bits, qubits, and qudits.

One thing of note is that the kinemetics are fixed but the realizations can be different. For example the classical $q=4$ theory and qubit theory both have the same MacWilliams transforms (since they both have $r=1$, $N=4$) but their realizations are very different. 

Now we have a family with a single parameter $N$. It's curious to take the limit as $N \to \infty$. For the standard theories with $r=1$, we have
\begin{equation}
\beta = 2\arccos(\tfrac{1}{\sqrt{N}}) \to 2\arccos(0) = \pi 
\end{equation}
So we obtain the Wigner D matrix
\[
\widehat U_N\longrightarrow -i  X
=
D^{1/2}\left(\frac{\pi}{2},\pi,\frac{\pi}{2}\right).
\]
Equivalently, $ U_N\longrightarrow
 X=
\begin{pmatrix}
0&1\\
1&0
\end{pmatrix} $. And thus at length \(n\),
\[
U_N^{(n)}\longrightarrow\Sym^n(X),
\]
the reversal matrix \(k\mapsto n-k\).  This is precisely Rains'
unitary-enumerator duality,
\[
A'_k=B'_{n-k},
\qquad
A'(x,y)=B'(y,x)
\]
\cite[Theorem~18]{rainsWE}.  Thus, at the level of duality
transformations, Rains' unitary enumerators realize the formal
large-\(N\) endpoint of the MacWilliams spin family.
In summary we can think of the MacWilliams transforms for classical bits, classical $q$-ary bits, qubits, qudits, and also Rains' unitary enumerators as part of the same contiguous family determined entirely by the parameter $N$ or equivalently the rotation angle $\beta = 2 \arccos( \frac{1}{\sqrt{N}})$. This is rather surprising because the classical and quantum theories are \textit{both} dictated by spin, which is a decidedly quantum phenomenon!

\emph{Qubit shadows are kinematically forced.--}
So far we have focused on the transformation between the local-sector basis $\{\ket{0}, \ket{1}\}$ and the global-sector basis $\{\ket{+},  \ket{-}\}$. However, recall there was a $\mathbb{Z}_2$ gauge choice for the latter basis since $\ket{-}$ was not the only canonical choice. In particular there is a second \textit{twisted global-sector} basis $\{ \ket{+}, \ket{\sigma}\}$ where $\ket{\sigma} = - \ket{-}$. This is just as canonical of a basis as the global-sector basis. And even though this twisted basis differs only by a seemingly innocuous sign from the global-sector basis, it is entirely responsible for the kinematics undergirding qubit quantum shadows.

Consider the change of basis matrix $S$ on one-site between the local-sector basis $\{\ket{0},\ket{1}\}$ and the twisted global basis $\{\ket{+}, \ket{\sigma}\}$. It is easy to see that
\[
    S = U Z, \qquad Z = \smqty(1 & 0 \\ 0 & -1).  \numberthis
\]
On $n$-sites we can symmetrize like before to get
\[
    S^{(n)}  = \Sym^n(S) = U^{(n)} Z^{(n)} \numberthis
\]
where $Z^{(n)} = \mathrm{diag}(1 , -1, 1, -1, \cdots)$. 

Because $Z$ and $\Delta$ are diagonal, they commute and thus the shadow transform in counting coordinates is
\begin{equation}
  M_{S}^{(n)}=M^{(n)}Z^{(n)},
  \qquad
  (M_{S}^{(n)})_{j,k}=(-1)^kM^{(n)}_{j,k}.
  \label{eq:counting-shadow}
\end{equation}

The Rains qubit shadow enumerator \cite{Shadows} is defined as
\[
   \frac{1}{K}\sum_{\mathrm{wt}(E)=j}\Tr( PEY^{\otimes n} P^* Y^{\otimes n} E^\dagger), \numberthis
\]
where $Y = \smqty(0 & -i \\ i & 0)$ and $P^*$ is the complex conjugate of the code projector. The shadow transformation is then
\begin{align*}
    \mathbf{x} &\to \mathbf{x} + 3\mathbf{y} \\
    \mathbf{y} &\to \mathbf{y}-\mathbf{x} = -(\mathbf{x}-\mathbf{y})
\end{align*}
Indeed this matches \cref{eq:counting-shadow} and so the change of basis between the local-sector basis and the twisted global-sector basis is the kinematics responsible for Rains' quantum shadow enumerators. So this is the conceptual payoff: for qubits, MacWilliams and shadow are the two orientations of the same canonical sector-frame change. 

Note this canonical shadow transform does \textit{not} produce the classical binary Conway--Sloane shadow, but for a very specific reason that we address in the appendix.

\emph{Heterogeneous enumerators.--} So far we have looked at $n$ identical sites. But instead suppose site $i$ has local dimension $q_i$, 
which may vary from site to site. We can still use our framework to write down the kinematical MacWilliams transforms in this case.

For ease, suppose each site has $r=1$ trivial error symbol (the identity) and $N_i = q_i^2$ total error symbols. So the one-site 
unitary MacWilliams transform at site $i$ is just
\[
    U_{i}
    =
    \frac{1}{q_i}
    \begin{pmatrix}
        1 & \sqrt{q_i^2-1} \\
        \sqrt{q_i^2-1} & -1
    \end{pmatrix}. \numberthis
\]
The full heterogeneous transform is just the tensor product of these one-site 
transforms, one per site:
\[
    U_{\mathrm{het}}
    =
    U_1 \otimes \cdots \otimes U_n. \numberthis
\]
If there happen to be $m_q$ sites of the same local dimension $q$, those 
sites contribute a spin-$m_q/2$ Wigner-$D$ block. Grouping equal dimensions 
together (this sort of grouping of certain sites is essentially the case of split weight enumerators \cite{SplitWE}) gives
\[
    U_{\mathrm{het}}
    =
    \bigotimes_q\,
    i^{m_q} D^{m_q/2}\left(\frac{\pi}{2},\beta_q,\frac{\pi}{2}\right), \qquad \beta_q = 2 \arccos(\tfrac{1}{q}).
\]
So the single Wigner-$D$ matrix of the homogeneous case is replaced by a 
tensor product of Wigner-$D$ matrices, one for each distinct local dimension. 
Each factor is the same construction as before, just at a different value of 
$N = q^2$.

Passing to counting coordinates gives the mixed-dimensional MacWilliams 
substitution independently for each local dimension $q$:
\[
    \mathbf{x}_q \longmapsto \mathbf{x}_q + (q^2-1)\mathbf{y}_q,
    \qquad
    \mathbf{y}_q \longmapsto \mathbf{x}_q - \mathbf{y}_q.
\]
This is exactly the mixed-dimensional quantum MacWilliams identity of 
Ref.~\cite[Theorem~12]{MixedDimMacWilliams} up to a different choice of normalization. Its generalized Krawtchouk form is obtained by the same symmetric-power
argument used for the homogeneous case in the appendix.

\emph{$\Gamma$-orbit weight enumerators.--}
Let $V$ be the two-sector space on one site. Ordinary weight enumerators live in
$$
\Sym^n(V)\cong\Inv_{S_n}(V^{\otimes n}),
$$
which forgets all positional information. More generally, for any subgroup $\Gamma\leq S_n$, one may retain the orbit-resolved information in
$$
\Inv_{\Gamma}(V^{\otimes n}).
$$
Since $U^{\otimes n}$ commutes with the permutation action on $V^{\otimes n}$, the one-site MacWilliams transform restricts canonically to each such invariant space.

For example, when $\Gamma=C_n$, the basis is indexed by cyclic support patterns, or necklaces. On six sites, the weight-two patterns $110000$, $101000$, and $100100$ describe adjacent errors, errors separated by one site, and errors separated by two sites, respectively. Such an enumerator distinguishes supports of equal weight but different spatial extent, and may therefore be useful for localized, burst, or correlated noise, for which error likelihood is not determined by weight alone. Mathematically, this produces an intermediate theory between full enumerators and the symmetrized ordinary enumerators. It would be an interesting line of future work to determine what additional constraints and coding bounds survive at this finer precision. 

\emph{Higher resolution.--}
The two-sector theory $\E_0 \sqcup \E_1$ above is only the lowest-resolution case. More generally, suppose the one-site label set is partitioned into $a$ sectors $\E_0 \sqcup \cdots \sqcup \E_{a-1}$, with associated sector space $V$ satisfying $\dim V=a$. If all sites are identical and their positions are forgotten, the natural $n$-site space is
$$
\Sym^n(V).
$$
This space carries the symmetric-power representation of $\U(a)$, or equivalently of $\SU(a)$ up to an overall phase. Thus any chosen one-site unitary $U$ induces a canonical $n$-site transform $\Sym^n(U)$, whose matrix elements are multinomial analogues of the ordinary Krawtchouk coefficients.

What is no longer automatic for $a>2$ is the choice of the one-site transform itself. In the two-sector case, the uniform direction has a one-dimensional mean-zero complement, so the global frame is fixed up to orientation and the MacWilliams and shadow transforms are essentially forced. For $a>2$, the mean-zero complement has dimension $a-1$, and the sector partition alone leaves a nontrivial freedom in choosing a global frame. A distinguished higher-rank MacWilliams transform therefore requires either an additional kinematic prescription or further algebraic structure supplied by the realization, such as a character table, a dual partition, or an association scheme.

Complete weight enumerators correspond formally to the extreme case $a=N$, in which every local label is retained as its own sector. Their familiar MacWilliams transforms arise not merely from this finest partition, but from the additional Fourier-duality structure of the local alphabet. It would be interesting to determine which sector decompositions admit similarly natural transforms and how their symmetric powers organize multivariate enumerator identities and higher-rank shadow theories.

\emph{Conclusion.--}
The MacWilliams transform is spin kinematics in disguise. Once a weight enumerator remembers only whether each site is inactive or active, its $n$-site space is forced to be the spin-$n/2$ representation of $\SU(2)$. The unitary MacWilliams transform is therefore a Wigner-$D$ matrix, and the familiar Krawtchouk matrix is nothing more than the same rotation written in counting coordinates.

At the level of kinematics, the distinction between classical and quantum disappears. What appeared to be a collection of only loosely related identities is therefore a single unified spin mechanism. This shifts the problem from deriving each transform separately to understanding which realizations, symmetries, and positivity structures can inhabit the common kinematical space.

\bibliography{biblio.bib}

\begin{thebibliography}{13}%
\makeatletter
\providecommand \@ifxundefined [1]{%
 \@ifx{#1\undefined}
}%
\providecommand \@ifnum [1]{%
 \ifnum #1\expandafter \@firstoftwo
 \else \expandafter \@secondoftwo
 \fi
}%
\providecommand \@ifx [1]{%
 \ifx #1\expandafter \@firstoftwo
 \else \expandafter \@secondoftwo
 \fi
}%
\providecommand \natexlab [1]{#1}%
\providecommand \enquote  [1]{``#1''}%
\providecommand \bibnamefont  [1]{#1}%
\providecommand \bibfnamefont [1]{#1}%
\providecommand \citenamefont [1]{#1}%
\providecommand \href@noop [0]{\@secondoftwo}%
\providecommand \href [0]{\begingroup \@sanitize@url \@href}%
\providecommand \@href[1]{\@@startlink{#1}\@@href}%
\providecommand \@@href[1]{\endgroup#1\@@endlink}%
\providecommand \@sanitize@url [0]{\catcode `\\12\catcode `\$12\catcode `\&12\catcode `\#12\catcode `\^12\catcode `\_12\catcode `\%12\relax}%
\providecommand \@@startlink[1]{}%
\providecommand \@@endlink[0]{}%
\providecommand \url  [0]{\begingroup\@sanitize@url \@url }%
\providecommand \@url [1]{\endgroup\@href {#1}{\urlprefix }}%
\providecommand \urlprefix  [0]{URL }%
\providecommand \Eprint [0]{\href }%
\providecommand \doibase [0]{https://doi.org/}%
\providecommand \selectlanguage [0]{\@gobble}%
\providecommand \bibinfo  [0]{\@secondoftwo}%
\providecommand \bibfield  [0]{\@secondoftwo}%
\providecommand \translation [1]{[#1]}%
\providecommand \BibitemOpen [0]{}%
\providecommand \bibitemStop [0]{}%
\providecommand \bibitemNoStop [0]{.\EOS\space}%
\providecommand \EOS [0]{\spacefactor3000\relax}%
\providecommand \BibitemShut  [1]{\csname bibitem#1\endcsname}%
\let\auto@bib@innerbib\@empty
\bibitem [{\citenamefont {MacWilliams}(1963)}]{MacWilliams1963}%
  \BibitemOpen
  \bibfield  {author} {\bibinfo {author} {\bibfnamefont {F.~J.}\ \bibnamefont {MacWilliams}},\ }\bibfield  {title} {\bibinfo {title} {A theorem on the distribution of weights in a systematic code},\ }\href@noop {} {\bibfield  {journal} {\bibinfo  {journal} {Bell System Technical Journal}\ }\textbf {\bibinfo {volume} {42}},\ \bibinfo {pages} {79} (\bibinfo {year} {1963})}\BibitemShut {NoStop}%
\bibitem [{\citenamefont {MacWilliams}\ and\ \citenamefont {Sloane}(1977)}]{MacWilliamsSloane1977}%
  \BibitemOpen
  \bibfield  {author} {\bibinfo {author} {\bibfnamefont {F.~J.}\ \bibnamefont {MacWilliams}}\ and\ \bibinfo {author} {\bibfnamefont {N.~J.~A.}\ \bibnamefont {Sloane}},\ }\href@noop {} {\emph {\bibinfo {title} {The Theory of Error-Correcting Codes}}}\ (\bibinfo  {publisher} {North-Holland},\ \bibinfo {year} {1977})\BibitemShut {NoStop}%
\bibitem [{\citenamefont {Delsarte}(1973)}]{Delsarte1973}%
  \BibitemOpen
  \bibfield  {author} {\bibinfo {author} {\bibfnamefont {P.}~\bibnamefont {Delsarte}},\ }\href@noop {} {\emph {\bibinfo {title} {An Algebraic Approach to the Association Schemes of Coding Theory}}},\ Vol.~\bibinfo {volume} {10}\ (\bibinfo  {publisher} {Philips Research Reports Supplements},\ \bibinfo {address} {Eindhoven},\ \bibinfo {year} {1973})\BibitemShut {NoStop}%
\bibitem [{\citenamefont {Conway}\ and\ \citenamefont {A.}(2008)}]{ConwaySloane}%
  \BibitemOpen
  \bibfield  {author} {\bibinfo {author} {\bibfnamefont {J.~H.}\ \bibnamefont {Conway}}\ and\ \bibinfo {author} {\bibfnamefont {S.~N.~J.}\ \bibnamefont {A.}},\ }\href@noop {} {\emph {\bibinfo {title} {Sphere packings, lattices, and groups}}}\ (\bibinfo  {publisher} {World Publishing Corp.},\ \bibinfo {year} {2008})\BibitemShut {NoStop}%
\bibitem [{\citenamefont {Rains}(1998)}]{rainsWE}%
  \BibitemOpen
  \bibfield  {author} {\bibinfo {author} {\bibfnamefont {E.}~\bibnamefont {Rains}},\ }\bibfield  {title} {\bibinfo {title} {Quantum weight enumerators},\ }\href {https://doi.org/10.1109/18.681316} {\bibfield  {journal} {\bibinfo  {journal} {IEEE Transactions on Information Theory}\ }\textbf {\bibinfo {volume} {44}},\ \bibinfo {pages} {1388} (\bibinfo {year} {1998})}\BibitemShut {NoStop}%
\bibitem [{\citenamefont {Rains}(1999)}]{Shadows}%
  \BibitemOpen
  \bibfield  {author} {\bibinfo {author} {\bibfnamefont {E.}~\bibnamefont {Rains}},\ }\bibfield  {title} {\bibinfo {title} {Quantum shadow enumerators},\ }\href {https://doi.org/10.1109/18.796376} {\bibfield  {journal} {\bibinfo  {journal} {IEEE Transactions on Information Theory}\ }\textbf {\bibinfo {volume} {45}},\ \bibinfo {pages} {2361} (\bibinfo {year} {1999})}\BibitemShut {NoStop}%
\bibitem [{\citenamefont {Shor}\ and\ \citenamefont {Laflamme}(1997)}]{ShorWE}%
  \BibitemOpen
  \bibfield  {author} {\bibinfo {author} {\bibfnamefont {P.}~\bibnamefont {Shor}}\ and\ \bibinfo {author} {\bibfnamefont {R.}~\bibnamefont {Laflamme}},\ }\bibfield  {title} {\bibinfo {title} {Quantum analog of the macwilliams identities for classical coding theory},\ }\href {https://doi.org/10.1103/PhysRevLett.78.1600} {\bibfield  {journal} {\bibinfo  {journal} {Phys. Rev. Lett.}\ }\textbf {\bibinfo {volume} {78}},\ \bibinfo {pages} {1600} (\bibinfo {year} {1997})}\BibitemShut {NoStop}%
\bibitem [{\citenamefont {Rains}(2000)}]{RainsPolys}%
  \BibitemOpen
  \bibfield  {author} {\bibinfo {author} {\bibfnamefont {E.}~\bibnamefont {Rains}},\ }\bibfield  {title} {\bibinfo {title} {Polynomial invariants of quantum codes},\ }\href {https://doi.org/10.1109/18.817508} {\bibfield  {journal} {\bibinfo  {journal} {IEEE Transactions on Information Theory}\ }\textbf {\bibinfo {volume} {46}},\ \bibinfo {pages} {54} (\bibinfo {year} {2000})}\BibitemShut {NoStop}%
\bibitem [{\citenamefont {Lai}\ and\ \citenamefont {Ashikhmin}(2018)}]{SplitWE}%
  \BibitemOpen
  \bibfield  {author} {\bibinfo {author} {\bibfnamefont {C.-Y.}\ \bibnamefont {Lai}}\ and\ \bibinfo {author} {\bibfnamefont {A.}~\bibnamefont {Ashikhmin}},\ }\bibfield  {title} {\bibinfo {title} {Linear programming bounds for entanglement-assisted quantum error-correcting codes by split weight enumerators},\ }\href {https://doi.org/10.1109/TIT.2017.2711601} {\bibfield  {journal} {\bibinfo  {journal} {IEEE Transactions on Information Theory}\ }\textbf {\bibinfo {volume} {64}},\ \bibinfo {pages} {622} (\bibinfo {year} {2018})}\BibitemShut {NoStop}%
\bibitem [{\citenamefont {Hu}\ \emph {et~al.}(2019)\citenamefont {Hu}, \citenamefont {Yang},\ and\ \citenamefont {Yau}}]{WEdoubleandcomplete}%
  \BibitemOpen
  \bibfield  {author} {\bibinfo {author} {\bibfnamefont {C.}~\bibnamefont {Hu}}, \bibinfo {author} {\bibfnamefont {S.}~\bibnamefont {Yang}},\ and\ \bibinfo {author} {\bibfnamefont {S.~S.-T.}\ \bibnamefont {Yau}},\ }\bibfield  {title} {\bibinfo {title} {Complete weight distributions and macwilliams identities for asymmetric quantum codes},\ }\href {https://doi.org/10.1109/ACCESS.2019.2918529} {\bibfield  {journal} {\bibinfo  {journal} {IEEE Access}\ }\textbf {\bibinfo {volume} {7}},\ \bibinfo {pages} {68404} (\bibinfo {year} {2019})}\BibitemShut {NoStop}%
\bibitem [{\citenamefont {Hu}\ \emph {et~al.}(2020)\citenamefont {Hu}, \citenamefont {Yang},\ and\ \citenamefont {Yau}}]{CompleteWeightEnumQuantum}%
  \BibitemOpen
  \bibfield  {author} {\bibinfo {author} {\bibfnamefont {C.}~\bibnamefont {Hu}}, \bibinfo {author} {\bibfnamefont {S.}~\bibnamefont {Yang}},\ and\ \bibinfo {author} {\bibfnamefont {S.~S.-T.}\ \bibnamefont {Yau}},\ }\bibfield  {title} {\bibinfo {title} {Weight enumerators for nonbinary asymmetric quantum codes and their applications},\ }\href@noop {} {\bibfield  {journal} {\bibinfo  {journal} {Advances in Applied Mathematics}\ }\textbf {\bibinfo {volume} {121}},\ \bibinfo {pages} {102085} (\bibinfo {year} {2020})}\BibitemShut {NoStop}%
\bibitem [{\citenamefont {Cao}\ and\ \citenamefont {Lackey}(2024)}]{WEtensor}%
  \BibitemOpen
  \bibfield  {author} {\bibinfo {author} {\bibfnamefont {C.}~\bibnamefont {Cao}}\ and\ \bibinfo {author} {\bibfnamefont {B.}~\bibnamefont {Lackey}},\ }\bibfield  {title} {\bibinfo {title} {Quantum weight enumerators and tensor networks},\ }\href {https://doi.org/10.1109/TIT.2023.3340503} {\bibfield  {journal} {\bibinfo  {journal} {IEEE Transactions on Information Theory}\ }\textbf {\bibinfo {volume} {70}},\ \bibinfo {pages} {3512} (\bibinfo {year} {2024})}\BibitemShut {NoStop}%
\bibitem [{\citenamefont {González-Lociga}\ and\ \citenamefont {Ball}(2026)}]{MixedDimMacWilliams}%
  \BibitemOpen
  \bibfield  {author} {\bibinfo {author} {\bibfnamefont {D.}~\bibnamefont {González-Lociga}}\ and\ \bibinfo {author} {\bibfnamefont {S.}~\bibnamefont {Ball}},\ }\href {https://arxiv.org/abs/2604.25790} {\bibinfo {title} {The mixed-dimensional quantum macwilliams identity: bounds for codes and absolutely maximally entangled states in heterogeneous systems}} (\bibinfo {year} {2026}),\ \Eprint {https://arxiv.org/abs/2604.25790} {arXiv:2604.25790 [quant-ph]} \BibitemShut {NoStop}%
\end{thebibliography}%

\clearpage
\appendix
\begin{center}
    {\Large Supplemental Material}
\end{center}

\section{Larger Trivial Sectors}
We have retained arbitrary sector sizes $r$ and $s$ because the construction requires only a partition of the one-site action alphabet into two classes. Although the standard coding-theoretic choice has $r=1$, with the identity as the unique trivial action, other effective descriptions may regard several distinct local actions as equivalent or inert relative to the information being protected. This could occur, for example, in subsystem, noiseless-subsystem, or other coarse-grained noise models. Whenever such a distinction produces two sectors, the same canonical MacWilliams and shadow kinematics continues to apply, with the transform determined solely by $r$ and $s$. It would be interesting to identify settings in which this more general notion of weight captures information not visible to the usual identity-versus-nonidentity enumerator.

\section{The \(ZXZ\) convention for Wigner matrices}

To fix phases and basis conventions, we use the \(ZXZ\) Euler-angle
parameterization
\begin{equation}
D^j(\alpha,\beta,\gamma)
:=
e^{-i\alpha J_z^{(j)}}
e^{-i\beta J_x^{(j)}}
e^{-i\gamma J_z^{(j)}} .
\label{eq:ZXZ-convention}
\end{equation}
This differs from the frequently used \(ZYZ\) convention.  In the spin-\(1/2\)
representation, with basis ordered as \(\{\ket{0},\ket{1}\}\), we take
\begin{equation}
J_z^{(1/2)}=\frac{ Z}{2},
\qquad
J_x^{(1/2)}=\frac{ X}{2}.
\end{equation}
Consequently,
\begin{align}
D^{1/2}\left(\frac{\pi}{2},\beta,\frac{\pi}{2}\right)
&=
e^{-i\pi Z/4}
e^{-i\beta X/2}
e^{-i\pi Z/4}
\nonumber\\
&=
-i
\begin{pmatrix}
\cos(\beta/2) & \sin(\beta/2)\\
\sin(\beta/2) & -\cos(\beta/2)
\end{pmatrix}.
\label{eq:ZXZ-spin-half}
\end{align}
For sector sizes \(r\) and \(s\), let \(N=r+s\) and choose
\begin{equation}
\beta
=
2\arccos\sqrt{\frac{r}{N}},
\qquad
0\leq\beta\leq\pi .
\end{equation}
Then
\begin{equation}
\cos\frac{\beta}{2}=\sqrt{\frac{r}{N}},
\qquad
\sin\frac{\beta}{2}=\sqrt{\frac{s}{N}},
\end{equation}
and hence
\begin{equation}
D^{1/2}\left(\frac{\pi}{2},\beta,\frac{\pi}{2}\right)
=
-iU,
\qquad
U=
\frac{1}{\sqrt N}
\begin{pmatrix}
\sqrt r&\sqrt s\\
\sqrt s&-\sqrt r
\end{pmatrix}.
\label{eq:ZXZ-MacWilliams}
\end{equation}
Thus our special-unitary representative of the one-site MacWilliams
transformation is \(\widehat U=-iU\).

Under the standard identification
\(\Sym^n(\CC^2)\cong\mathcal H_{n/2}\), with the weight-\(k\) vector
corresponding to magnetic quantum number \(m=n/2-k\), the same convention
gives
\begin{equation}
\Sym^n(\widehat U)
=
D^{n/2}\left(\frac{\pi}{2},\beta,\frac{\pi}{2}\right).
\end{equation}
Since \(\widehat U=-iU\), the phase relating the physical orthogonal
MacWilliams transformation to the Wigner matrix is therefore
\begin{equation}
U^{(n)}
=
\Sym^n(U)
=
i^nD^{n/2}\left(\frac{\pi}{2},\beta,\frac{\pi}{2}\right).
\label{eq:ZXZ-many-site-phase}
\end{equation}
Equations~\eqref{eq:ZXZ-convention}--\eqref{eq:ZXZ-many-site-phase} fix all
Euler-angle, basis-ordering, and overall-phase conventions used in the main
text.

\section{Deriving the Krawtchouk Formula}

The one-site MacWilliams transform for arbitrary inactive and active sector
sizes is
$$
M=
\mqty(
r & r \\
s & -r
).
$$
It acts on the counting basis $\{\mathbf{x},\mathbf{y}\}$ as
$$
M\mathbf{x}=r\mathbf{x}+s\mathbf{y},
\qquad
M\mathbf{y}=r\mathbf{x}-r\mathbf{y}.
$$
The corresponding $n$-site MacWilliams transform is the symmetric power
$$
M^{(n)}=\operatorname{Sym}^n(M).
$$

The space $\operatorname{Sym}^n(V)$ can be realized as the space of
homogeneous polynomials of degree $n$ in $\mathbf{x}$ and $\mathbf{y}$. A
natural weight basis is
$$
\mathbf{e}_k=\mathbf{x}^{n-k}\mathbf{y}^k,
\qquad
k=0,1,\ldots,n.
$$
The basis vector $\mathbf{e}_k$ represents a configuration with $k$ active
sites and $n-k$ inactive sites.

The symmetric-power action is obtained by applying the one-site
transformation to each factor. Therefore,
$$
M^{(n)}\mathbf{e}_k
=
(M\mathbf{x})^{n-k}(M\mathbf{y})^k,
$$
and hence
$$
M^{(n)}\mathbf{e}_k
=
(r\mathbf{x}+s\mathbf{y})^{n-k}
(r\mathbf{x}-r\mathbf{y})^k.
$$

To obtain the matrix element $M^{(n)}_{j,k}$, we extract the coefficient of
the output weight-$j$ basis vector
$$
\mathbf{e}_j=\mathbf{x}^{n-j}\mathbf{y}^j.
$$
Such a term contains exactly $j$ powers of $\mathbf{y}$. Suppose that
$\ell$ of them are chosen from the second factor
$(r\mathbf{x}-r\mathbf{y})^k$. There are $\binom{k}{\ell}$ ways to make
this choice, and these terms contribute $(-r)^\ell$. The remaining
$j-\ell$ powers of $\mathbf{y}$ must then be chosen from the first factor
$(r\mathbf{x}+s\mathbf{y})^{n-k}$. There are
$\binom{n-k}{j-\ell}$ ways to do so, and these terms contribute
$s^{j-\ell}$. 

After choosing the $j$ factors that contribute $\mathbf{y}$, all remaining factors must contribute their $\mathbf{x}$ term. Since the $\mathbf{x}$ term is $r\mathbf{x}$ in both binomials, these $n-j$ factors contribute $r^{n-j}$.

Multiplying these contributions and summing over all possible values of
$\ell$ gives
$$
M^{(n)}_{j,k}
=
\sum_{\ell=0}^{j}
(-1)^\ell
\binom{k}{\ell}
\binom{n-k}{j-\ell}
r^{n-j+\ell}s^{j-\ell}.
$$
Here binomial coefficients outside their natural range are understood to
vanish. Thus the matrix elements of $M^{(n)}$ are precisely the
Krawtchouk coefficients associated with the sector sizes $r$ and $s$.

\section{Classical shadows and complex gauge choices}

The transform $M^{(n)}Z^{(n)}$ between the local-sector basis and the twisted global-sector basis in the main text corresponded exactly to qubit quantum shadows. However, it is not responsible for the binary Conway--Sloane classical shadow. The reason is that the Conway--Sloane shadow is applicable only to a restricted class of binary classical codes, namely, \textit{even} classical codes where every codeword has even weight. Consequently, $Z^{(n)}$ acts trivially on these codes and so the transform $M^{(n)}Z^{(n)}$ acts like the MacWilliams transform $M^{(n)}$ and thus adds no new information. 

The idea is that $Z^{(n)}$ distinguishes between codewords of weight-$0$ and weight-$1$ (modulo 2). If the code is even then all codewords are weight-$0$ (modulo-2). So the binary Conway--Sloane classical shadow instead distinguishes between weight-$0$ and weight-$2$ (modulo 4). Thus instead of using $Z^{(n)}$ we need to use $T_{\pi/2}^{(n)}$ where $T_{\pi/2} = \smqty(1 & 0 \\ 0 & i)$ is the quantum phase gate. 

Thus the Conway--Sloane shadow transform corresponds to
\[
M^{(n)}T_{\pi/2}^{(n)}.
\]
But all is not lost: this transform \textit{can} be incorporated into the present framework by complexifying the one-site sector space. Over $\R$, the mean-zero $G$-invariant global line has only the two orientations $\ket{-}$ and $\ket{\sigma} = -\ket{-}$, giving the $\mathbb Z_2$ gauge choice of the main text. But over $\CC$, this enlarges to a $\U(1)$ gauge freedom corresponding to
$$
 e^{i\theta}\ket{-}.
$$
The change of basis matrix between the local sector basis $\{ \ket{0} , \ket{1} \}$ and the $U(1)$-twisted global-basis $\{ \ket{+}, e^{i \theta} \ket{-} \}$ is a \textit{$\theta$-twisted MacWilliams} transform
$$
M^{(n)}T_\theta^{(n)},
\qquad
T_\theta=\mqty(1&0 \\ 0&e^{i\theta}),
$$
where the weight-$k$ sector acquires the phase $e^{ik\theta}$. The choices
$$
\theta=0,\qquad
\theta=\pi,\qquad
\theta=\frac{\pi}{2}
$$
give the ordinary MacWilliams transform, the qubit quantum shadow transform of the main text, and the binary classical Conway--Sloane transform, respectively.

But unlike the real $\mathbb Z_2$ choice, this $\U(1)$ extension is not particularly canonical: there are infinitely many possible choices of phase, and a generic choice produces complex coefficients with no linear-programming interpretation. Particular phases become useful only after restricting the structure of the code.

It would be interesting to develop this hierarchy systematically. The kinematics supplies the entire family of $\theta$-twisted MacWilliams transforms; the remaining problem is to determine which classical or quantum realizations make a chosen twist real, nonnegative, and operationally meaningful. This may lead to new shadow theories for structured $q$-ary codes, as well as refined enumerators for even or more highly divisible quantum codes. Such conditions are also naturally related to the divisibility and phase constraints that appear in transversal $T$ gates and more general transversal diagonal gates.

\end{document}